\documentclass[a4paper,11pt]{article}

\usepackage[top=1in, bottom=1in, left=1in, right=1in]{geometry}

\usepackage{mcode}
\usepackage{diagbox}
\usepackage{algorithm}
\usepackage{algpseudocode}
\usepackage{array}
\usepackage{url}
\usepackage{bm}
\usepackage{graphicx}
\usepackage{color}

\usepackage[caption=false]{subfig}
\usepackage{hyperref}
\usepackage{datetime}

\newcommand{\old}{\text{old}}

\newcommand{\CSI}{\text{CSI}}
\newcommand{\CCCSI}{\text{CC-CSI}}
\newcommand{\MRCSI}{\text{MR-CSI}}

\newcommand{\sct}{\text{sct}}
\newcommand{\inc}{\text{inc}}

\graphicspath{{figures/}}

\begin{document}

\title{Cross-correlated Contrast Source Inversion}

\author{
        Shilong~Sun\thanks{S. Sun, B. J. Kooij, and A. G. Yarovoy are with the Delft University of Technology, 2628 Delft,
        The Netherlands (e-mail: S.Sun@tudelft.nl; B.J.Kooij@tudelft.nl; A.Yarovoy@tudelft.nl).}
        \and Bert~Jan~Kooij\footnotemark[1]
        \and Tian~Jin\thanks{T. Jin is with the College of Electronic Science and Engineering,
        National University of Defense Technology, Changsha 410073, China (e-mail: tianjin@nudt.edu.cn).}
        \and Alexander~G.~Yarovoy\footnotemark[1]}

\newdate{date}{20}{2}{2017}
\date{\displaydate{date}}

\maketitle

\begin{abstract}
In this paper, we improved the performance of the contrast source inversion (CSI) method by incorporating a so-called cross-correlated cost functional, which interrelates the state error and the data error in the measurement domain. The proposed method is referred to as the cross-correlated CSI. It enables better robustness and higher inversion accuracy than both the classical CSI and multiplicative regularized CSI (MR-CSI). In addition, we show how the gradient of the modified cost functional can be calculated without significantly increasing the computational burden. The advantages of the proposed algorithms are demonstrated using a 2-D benchmark problem excited by a transverse magnetic wave as well as a transverse electric wave, respectively, in comparison to classical CSI and MR-CSI.
\end{abstract}

\section{Introduction}\label{sec.Introduction}

  Inversion techniques have been applied extensively in many fields, e.g. radar imaging \cite{klotzsche2015crosshole}, seismic imaging \cite{hu2009simultaneous}, medical imaging \cite{rosenthal2013acoustic,gilmore2009microwave}, and so forth. Developments in inversion techniques and research are focused on computational efficiency, the incorporation of \textit{a priori} information to circumvent computational artifacts, and the calibration to the real antenna radiating pattern especially in near-field scenarios \cite{Serhir2008,Serhir2010,nounouh2015quantitative}. Methods to solve the inverse scattering problems include non-iterative methods, e.g. linear sampling method \cite{colton1997simple,colton1996simple}, and iterative methods \cite{van1997contrast}. The Contrast Source Inversion (CSI) method is an iterative frequency domain inversion method to retrieve the value of the contrast of scattering objects, which was first proposed by van den Berg et al. \cite{van1997contrast}, and was later applied to subsurface object detection in combination with integral equations based on the Electric Field Integral Equation (EFIE) formulation, see Kooij et al. \cite{kooij1999nonlinear}. \textit{A priori} information was introduced in the form of mathematical regularization constraints like the positivity constraints of the material properties and the Total Variation (TV) constraint in \cite{van1999extended} to further enhance the performance. In \cite{van1999extended}, a multiplicative regularized CSI (MR-CSI) method is proposed, in which the estimation of the tuning parameter is avoided. In addition, Crocco et al. \cite{isernia2004new,crocco2009contrast}, applied the so-called contrast source-extended Born-model to 2-D subsurface scattering problems. Later the CSI technique was introduced in combination with the finite-difference frequency domain (FDFD) scheme by Abubakar et al. \cite{abubakar2008finite,abubakar2011three}. The scheme based on the FDFD technique turned out to have computational advantages compared to EFIE scheme, especially if a non-homogeneous background, like the half-space configuration in ground penetrating radar (GPR), is required in the inversion. For a more accurate representation of complex geometry, finite-element method (FEM) was introduced and combined with CSI by Zakaria et al., and the 2-D inversion results with the transverse magnetic (TM) wave and the transverse electric (TE) wave can be found in \cite{zakaria2010finite} and \cite{zakaria2012finite}, respectively. FEM was applied as well to the contrast source-extended Born method in \cite{attardo2014contrast}. 

  One obvious drawback of the iterative methods is that a good initial guess must be provided beforehand to ensure the iterative inverting process converges to the global optimal solution. The reason is that there is more than one variable needed to be estimated during the inverting process. More specifically, the contrast and the contrast source are both unknown, and a less accurate initial guess is more likely to give a false gradient and thus leads the iterative inverting process to a local optimal solution. To overcome this drawback, the hybrid inversion schemes have been considered, which first recovered the shape of the scatterers faithfully by sampling-type technique, and then estimated their dielectric properties or improved the shape iteratively. The same idea can be found in the recently published papers \cite{crocco2012linear,ito2013two,kuo2014recursive,6867357}. 

  For iterative inversion methods, a cost function is normally needed which consists of the data error and the state error \cite{colton2012inverse}. On one hand, it ensures that the algorithm fits the measurement data. On the other hand, it tends to optimize the estimation of the contrast and the contrast source to satisfy the Maxwell equations. In this paper, we show that a minor state error can become large when mapped into the measurement domain because of the ill-posedness of the inverse scattering problem. In another word, there might be cases in which both the state error and data error are minimized sufficiently, while the state error is still large when mapped into the measurement domain, which indicates that the estimated contrast is not the global optimal solution. Inspired by this fact, we introduce a new error equation that interrelates the state error and the data error in the measurement domain and modifies the cost functional accordingly. In doing so, the state error and the data error are cross-correlated, and the inverting process is stabilized by minimizing the state error not only in the field domain, but also in the measurement domain. We refer to the proposed algorithm as cross-correlated contrast source inversion (CC-CSI) method. In addition, we also show how the gradient of the new cost functional can be calculated without significantly increasing the computational complexity. The performance of the proposed method is investigated based on a 2-D benchmark problem excited by a TM-polarized wave and a TE-polarized wave, respectively. As can be observed from the results, the CC-CSI method shows better robustness and higher inversion accuracy than classical CSI and MR-CSI. Since the Maxwell's equations are formulated in a three-dimensional FDFD formulation, it is already applicable to the reconstruction of future 3-D scattering objects. 

  The remainder of the paper is organized as follows: Section~\ref{sec.ProStaClaCSI} gives the problem statement and the introduction of classical CSI and MR-CSI. The cross-correlated CSI method is introduced in Section~\ref{sec.CroCorCSI}. Simulation results based on a 2-D benchmark problem are given in Section~\ref{sec.SimRes}, in which the performance investigation of CC-CSI in comparison to classical CSI and MR-CSI is fully discussed\footnote{The CC-CSI package is available at \url{https://github.com/TUDsun/CC-CSI}, in which MR-CSI and CSI are also contained for comparison.}. Finally, we give our conclusions in Section~\ref{sec.conclusion}. 

\section{Problem statement and Classical CSI Based on FDFD}\label{sec.ProStaClaCSI}

\subsection{Problem statement}\label{subsec.ProSta}
\begin{figure}[!ht]
  \centering
    \includegraphics[width=0.40\linewidth]{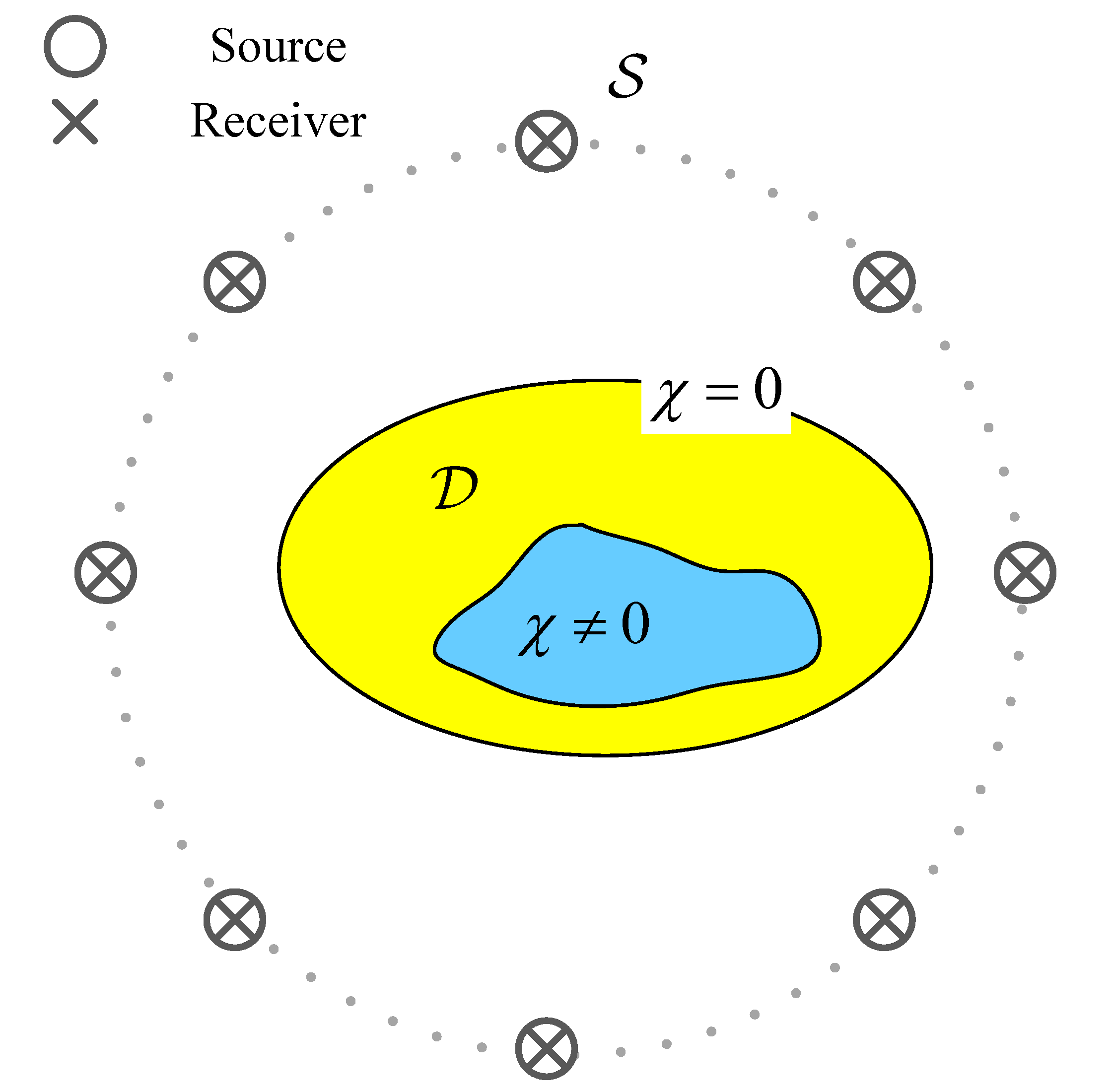}
    \caption{The configuration of the inverse scattering problem.}
    \label{fig:geo.CSI}
  \end{figure}

  We consider a scattering configuration as depicted in Figure~\ref{fig:geo.CSI}, 
  which consists of a bounded, simply connected, inhomogeneous background domain $\mcD$, which is also referred to as field domain in this paper. The domain $\mcD$ contains an object, whose location and complex permittivity profile are unknown. The measurement domain $\mcS$ contains the sources and receivers. The sources are denoted by the subscript $p$ in which $p\in\{1,2,3...,P\}$, the receivers are denoted by the subscript $q$ in which $q\in\{1,2,3,...,Q\}$. Sources and receivers that have equal subscripts are located at the same position. We use a right-handed coordinate system in which the unit vector in the invariant direction points out of the paper. The time factor of $\rme^{\rmi\omega t}$ is considered in this paper, where $\rmi^2=-1$.

  In our notation for the vectorial quantities, we use a bold notation which represents a vector with three components. The general mathematical representations presented are consistent with any 3-D configuration, in which the 2-D TE and TM excitations are a special case, resulting in vectors containing zero elements. According to the linear relation among the incident electric field $\mE^{\inc}_p$, the scattered electric field $\mE^{\sct}_p$ and the total electric fields $\mE_p$, $\mE^{\sct}_p=\mE_p-\mE^{\inc}_p$, the scattering equation with respect to the scattered electric field $\mE^{\sct}_p$ can be easily obtained which is \cite{sun2015TEandTM}
  \begin{equation}
    \nabla\times\vmu^{-1}\nabla\times\mE^{\sct}_p-\omega^2\vepsilon_b\mE^{\sct}_p=\omega^2\vchi\mE_p,
  \end{equation}
  with $p = 1,2,\dots,P$. Here, $\vmu$ represents the permeability of the background and is assumed to be equal to the permeability of the free space in this paper; $\omega$ is the angular frequency; the permittivity $\vvarepsilon_b$ and the conductivity $\vsigma_b$ of the background are incorporated into the complex permittivity $\vepsilon_b$ satisfying $\vepsilon_b=\vvarepsilon_b-\rmi\vsigma_b/\omega$. 

  The scattering equation can be further formulated based on FDFD scheme by \cite{sun2015TEandTM}
  \begin{equation}\label{eq.FD-CSI.eq}
    \mA\ve_p^{\sct}=\omega^2\vchi \ve_p,\quad p = 1,2,\dots,P,
  \end{equation}
  where $\mA$ is the FDFD stiffness matrix; $\ve_p^{\sct}$ and $\ve_p$ are the scattered electric field and the total electric field in the form of a column vector, respectively; and $\vchi$ is the contrast consisting of the difference of the permittivity $\vvarepsilon_\rmc$ and the difference of the conductivity $\vsigma_\rmc$ with the relation of $\vchi=\vvarepsilon_\rmc-\rmi\vsigma_\rmc/\omega$. Then the solution of \eqref{eq.FD-CSI.eq} is obtained by inverting the stiffness matrix $\mA$, which yields $\ve_p^{\sct}=\mA^{-1}\omega^2\vchi \ve_p$. This leads to the data equation
  \begin{equation}
    \vf_p=\mcM_\mcS \mA^{-1}\omega^2 \vchi \ve_p,\quad\vx\in\mcS,\quad p = 1,2,\dots,P,
  \end{equation}
  where $\mcM_{\mcS}$ is an operator that interpolates field values defined at the finite-difference grid points to the appropriate receiver positions. 

  In the remainder of this paper, $\omega^2$ is incorporated into $\mA$ for the sake of conciseness. The inverse scattering problem is to reconstruct the contrast $\vchi$ as a function of space from the incomplete measured field data $\{\vf_p\},\ p=1,2,\cdots, P$, which is full of challenges because of the nonlinearity and the ill-posedness.

\subsection{Classical CSI and MR-CSI}\label{subsec.ClaCSI}

  Classical CSI is a method of iteratively minimizing a cost functional consisting of the data error and the state error for reconstructing the contrast source. The contrast is updated during the iterations. Specifically, the multiplication of the contrast $\vchi$ and the total field $\ve_p$ is referred to as the contrast source, which is represented by $\vj_p$. Then, the data error $\vrho_p$ and the state error $\vgamma_p$ are defined by 
  \begin{eqnarray}
    \vrho_p=\vf_p-\mcM_{\mcS} \mA^{-1}\vj_p,\quad p = 1,2,\dots,P,
    \label{eq.object.error}
  \end{eqnarray}
  and
  \begin{equation}\label{eq.ep}
    \vgamma_p=\vchi \ve_p^{\inc}+\vchi\mcM_{\mcD}\mA^{-1}\vj_p-\vj_p,\quad p = 1,2,\dots,P,
  \end{equation}
  respectively. Here, $\mcM_{\mcD}$ is an operator that selects fields only inside the field domain $\mcD$. The cost functional is given by
  \begin{equation}\label{eq.CSIJCost}
    C_{\CSI}^{\vj}(\vchi_{n-1},\vj_p) = \eta^\mcS\sum_{p=1}^P\|\vrho_p\|^2_\mcS+\eta^\mcD\sum_{p=1}^P\|\vgamma_p\|^2_\mcD.
  \end{equation}
  with $\vchi = \vchi_{n-1}$. Here, $1/\eta^\mcS=\sum_p\|\vf_p\|^2_\mcS$, $1/\eta^\mcD=\sum_p\|\vchi \ve_p^{\inc}\|^2_\mcD$, and $\|\cdot\|_\mcS$ and $\|\cdot\|_\mcD$ represent the norms on the measurement space $L^2(\mcS)$ and the field space $L^2(\mcD)$, respectively. The contrast source $\vj_p$ is iteratively optimized by minimizing the cost functional $C^{\vj}(\vchi, \vj_p)$, followed by the update of the contrast $\vchi$ which is done by minimizing the cost functional
  \begin{equation}\label{eq.CSIChiCost}
    C_{\CSI}^{\vchi}(\vchi,\vj_{p,n}) = \left.\eta^\mcD\sum_{p=1}^P\|\vgamma_p\|^2_\mcD\right|_{\vj_p=\vj_{p,n}}.
  \end{equation}
  
  MR-CSI is the CSI method regularized with a multiplicative weighted total variation (TV) constraint, which was first introduced by van den Berg et al. \cite{van1999extended}. In comparison to CSI, the contrast $\vchi$ in MR-CSI is now updated by minimizing the cost functional
  \begin{equation}
    C_{\MRCSI}^{\vchi}(\vchi, \vj_{p,n}) =  \left.C_{\CSI}^{\vj}(\vchi, \vj_{p}) \times\frac{1}{V}\int_{\mcD}\frac{\|\nabla\vchi\|^2+\delta_{n-1}^2}{\|\nabla\vchi_{n-1}\|^2+\delta_{n-1}^2}d\vx\right|_{\vj_p=\vj_{p,n}} .
  \end{equation} 
  Here, $V=\int_\mcD d\vx$, and $\delta^2_{n-1}$ are introduced for restoring the differentiability of the TV factor \cite{van1999extended}. The value of $\delta^2_{n-1}$ is chosen to be large in the beginning of the optimization and small towards the end, which is given by 
  \begin{equation}
    \delta^2_{n-1} = C_{\CSI}^{\vchi}(\vchi_{n-1},\vj_{p,n})\Delta^2,
  \end{equation}
  where $\Delta$ denotes the mesh size of the discretized domain $\mcD$. It is worth noting that the contrast $\vchi$ is assumed to be isotropic in this paper. Namely, for the TE case, only one component of $\vchi$ is used in the TV penalty function. More details of classical CSI and MR-CSI can be found in \cite{abubakar2008finite,sun2015TEandTM,van1999extended,van2001contrast}.

\section{Cross-Correlated CSI}\label{sec.CroCorCSI}
  
\subsection{Motivation}\label{subsec.Motivation}

  As aforementioned, a good initial guess is very critical for ensuring that the iterative methods can successfully converge to the global optimal solution. This can be explained firstly by the fact that there are two unknown variables --- the contrast and the contrast source. The less accurate the initial guess is, the more inaccurate the gradient with respect to the contrast source will be. Secondly, although classical CSI is able to minimize the data error $\vrho(\vj_p)$ by constraining the state error $\vgamma(\vchi,\vj_p)$ at the same time, a global optimal solution can still not be guaranteed because of the ill-posedness of the inverse scattering problem. For simplicity, let us first define the measurement matrix as $\bm{\Phi}:=\mcM_\mcS \mA^{-1}$. The condition number of matrix $\bm{\Phi}$ is further defined as
  \begin{equation}
    \kappa(\bm{\Phi}):= \frac{\sigma_{\max}(\bm{\Phi})}{\sigma_{\min}(\bm{\Phi})}.
  \end{equation}
  where $\sigma_{\max}(\bm{\Phi})$ and $\sigma_{\min}(\bm{\Phi})$ are maximal and minimal singular values of $\bm{\Phi}$, respectively. As discussed in Subsection~\ref{subsec.CondPhi}, the measurement matrix has a large condition number, which means a minor state error in the field space $L^2(\mcD)$ may cause a large error in the measurement space $L^2(\mcS)$. This potential mismatch cannot be reflected by the cost functional of the classical CSI method. Inspired by this fact, we came up with the idea of introducing the so-called cross-correlated cost functional. 

\subsection{Cross-correlated CSI}\label{subsec.CroCorCSI}

  In this subsection, a new cost functional is proposed, which interrelates the mismatch of the state equation and the data error in the measurement space. This proposed algorithm is referred to as the cross-correlated contrast source inversion method. 

  Specifically, the cross-correlated error $\vxi_p$ is defined as
  \begin{equation}
    \vxi_p = \vf_p-\bm{\Phi}(\vchi \ve_p^{\inc}+\vchi\mcM_{\mcD}\mA^{-1}\vj_p).
  \end{equation}
  Note that if the state error $\vgamma_p$ is zero, then theoretically we have $\vrho_p = \vxi_p$. In classical CSI, sufficiently minimizing the cost function of Eq.~\eqref{eq.CSIJCost} does not necessarily mean that the cross-correlated error is sufficiently minimized. Therefore, the cost functional of the contrast source in the proposed CC-CSI method is modified and defined as
  \begin{equation}\label{eq.Cvj}
      C_\CCCSI^{\vj}(\vchi_{n-1},\vj_p) = \eta^\mcS\sum_{p=1}^P\|\vrho_p\|^2_\mcS + \left.\eta^\mcD\sum_{p=1}^P\|\vgamma_p\|^2_\mcD+\eta^\mcS\sum_{p=1}^P\|\vxi_p\|^2_\mcS\right|_{\vchi = \vchi_{n-1}}.
  \end{equation}
  Subsequently, the gradient (Fr{\'e}chet derivative) of the modified cost functional with respect to the contrast source $\vj_p$ becomes
  \begin{equation}
    \vg_{p,n} = -2\eta^\mcS\bm{\Phi}^H\vrho_p+2\eta^\mcD\left(\vchi\mcM_{\mcD}\mA^{-1}-\mI\right)^H\vgamma_p - \left.2\eta^\mcS\left(\bm{\Phi}\vchi\mcM_{\mcD}\mA^{-1}\right)^H\vxi_p\right|_{\vchi = \vchi_{n-1},\vj_p = \vj_{p,n-1}}.
  \end{equation}
  Here, $\mI$ represents the identity matrix, and $(\cdot)^H$ is the conjugate transpose operator. Now suppose $\vj_{p,n-1}$ and $\vchi_{n-1}$ are known, then we update $\vj_p$ by 
  \begin{equation}
    \vj_{p,n} = \vj_{p,n-1} + \alpha_{p,n}\vnu_{p,n},
  \end{equation}
  where $\alpha_{p,n}$ is constant and the update directions $\vnu_{p,n}$ are functions of the position. The update directions are chosen to be the Polak-Ribi{\`e}re conjugate gradient directions, which are given by
  \begin{equation}
    \begin{split}
      \vnu_{p,0} &= 0,\\
      \vnu_{p,n} &= \vg_{p,n}+\frac
                  {
                    \sum_{p'}
                    \left\langle 
                      \vg_{p',n},\vg_{p',n}-\vg_{p',n-1}
                    \right\rangle
                    _{\mcD}
                  }
                  {
                    \sum_{p'}
                      \left\|
                        \vg_{p',n-1}
                      \right\|
                      _{\mcD}^2
                  }
                  \vnu_{p,n-1},\quad n\geq1,
    \end{split}
  \end{equation}
  where $\left\langle\cdot\right\rangle_{\mcD}$ represents the inner product defined in the field space $L^2(\mcD)$. The step size $\alpha_{p,n}$ can be explicitly found by minimizing the cost functional $C_\CCCSI^{\vj}(\vchi_{n-1}, \vj_{p,n-1} + \alpha_{p,n}\vnu_{p,n})$ (see Appendix~\ref{sec.apdixA} for the derivation).

  Once the contrast source $\vj_{p,n}$ is determined, we update the contrast $\vchi$ by minimizing the cost functional of the contrast which is defined by
  \begin{equation}
    C_\CCCSI^{\vchi}(\vchi,\vj_{p,n}) = \eta^\mcD\sum_{p=1}^P\|\vgamma_p\|^2_\mcD + \eta^\mcS\sum_{p=1}^P\|\vxi_p\|^2_\mcS
  \end{equation}
  with $\vj_p=\vj_{p,n}$. Specifically, $\vchi$ is updated via 
  \begin{equation}\label{eq.chiupdating}
    \vchi_{n} = \vchi_{n-1} + \beta_{n}\vnu_{\vchi,n},
  \end{equation}
  where $\beta_n$ is constant and the update directions $\vnu_{p,n}$ are chosen to be the Polak-Ribi{\`e}re conjugate gradient directions, which are given by
  \begin{equation}\label{eq.chinu}
    \begin{split}
      \vnu_{\vchi,0} &= 0\\
      \vnu_{\vchi,n} &= \vg_{\vchi,n}+\frac
                  {
                    \left\langle 
                      \vg_{\vchi,n},\vg_{\vchi,n}-\vg_{\vchi,n-1}
                    \right\rangle
                    _{\mcD}
                  }
                  {
                      \left\|
                        \vg_{\vchi,n-1}
                      \right\|
                      _{\mcD}^2
                  }
                  \vnu_{\vchi,n-1}\ n\geq1,
    \end{split}
  \end{equation}
  where $\vg_{\vchi,n}$ is the preconditioned gradient of the contrast cost functional $C_\CCCSI^{\vchi}(\vchi)$ defined as
  \begin{equation}\label{eq.gchi}
    \vg_{\vchi,n} = \frac{2\eta^{\mcD}\sum_{p=1}^P\overline{\ve_{p,n}}\vgamma_{p}-2\eta^{\mcS}\sum_{p=1}^P\overline{\ve_{p,n}}\bm{\Phi}^H\vxi_{p}}{\sum_{p=1}^P\left|\ve_{p,n}\right|^2},
  \end{equation}
  with $\vchi=\vchi_{n-1}$, $\vj_p = \vj_{p,n}$, where $\overline{(\cdot)}$ represents the conjugate operator. The step size $\beta_n$ is determined by minimizing the cost function 
  \begin{equation}\label{eq.chicost}
    C_{\CCCSI,n}^{\vchi}=\frac{\displaystyle\sum_{p=1}^P\left\|(\vchi_{n-1}+\beta_n\vnu_{\vchi,n})\ve_{p,n}-\vj_{p,n}\right\|^2_{\mcD}}{\displaystyle\sum_{p=1}^P\left\|(\vchi_{n-1}+\beta_n\vnu_{\vchi,n})\ve^{\inc}_{p}\right\|^2_{\mcD}} + \eta^{\mcS}\sum_{p=1}^P\left\|\vf_p-\bm{\Phi}(\vchi_{n-1}+\beta_n\vnu_{\vchi,n})\ve_{p,n}\right\|^2_{\mcS}.
  \end{equation}
  This is a problem of finding the minimum of a single-variable function, and can be solved efficiently by the Brent's method \cite{brent1973algorithms,Forsythe1976computer}.

  \begin{algorithm}[ht]
    \caption{CC-CSI}\label{euclid}
    \begin{algorithmic}[1]
      \State Initialize $\vj_p$ 

      \State $\ve^{\sct}_p \gets \mA^{-1} \vj_p,\ \ve_p \gets \ve^{\sct}_p+\ve^{\inc}_p$

      \State $\eta^S \gets \left(\sum_{p=1}^P\left\|\vf_p\right\|_\mcS^2\right)^{-1}$

      \State $\vnu_p \gets \bm{0},\ \vnu_{\vchi} \gets \bm{0}$

      \State $\vchi \gets \frac{\sum_{p=1}^P\vj_p\overline{\ve_p}}{\sum_{p=1}^P\ve_p\overline{\ve_p}}$

      \While{$C(\vchi,\vj_p)<\delta$}

        \State $\eta^\mcD \gets \left(\sum_{p=1}^P\|\vchi \ve^{\inc}_p\|^2_\mcD\right)^{-1}$

            \State $\vrho_p \gets \vf_p-\bm{\Phi}\vj_p$

            \State $\vxi_p \gets \vf_p-\bm{\Phi}\vchi\ve_p$

            \State $\vgamma_p \gets \vchi \ve_p - \vj_p$

            \State $\vg_p^\old \gets \vg_p$

        \State $\vg_p \gets \mA^{-H}\overline{\vchi}(\eta^\mcD \vgamma_p-\eta^{\mcS} \bm{\Phi}^{H}\vxi_p )-\eta^{\mcS} \bm{\Phi}^{H}\vrho_p - \eta^\mcD\vgamma_p$

        \State $\vnu_p\gets \vg_p+\frac
                  {
                    \sum_{p'}
                    \left\langle 
                      \vg_{p'},\vg_{p'}-\vg_{p'}^\old
                    \right\rangle
                    _{\mcD}
                  }
                  {
                    \sum_{p'}
                      \left\|
                        \vg_{p'}^\old
                      \right\|
                      _{\mcD}^2
                  }
                  \vnu_p$

        \State $\ve_p^{\nu} \gets \mA^{-1}\vnu_p$

        \State $\alpha_p \gets -\frac{
                  \Re\left\{\left\langle \vg_p,\vnu_p\right\rangle_\mcD\right\}
              }{
                \eta^{\mcS}\left(\left\|\bm{\Phi}\vnu_p\right\|_{\mcS}^2+
                \left\|\bm{\Phi}\vchi\ve_p^\nu\right\|_{\mcS}^2\right)+
                \eta^\mcD\|\vnu_p-\vchi \ve_p^\nu\|_\mcD^2
              }$

        \State $\vj_p \gets \vj_p+\alpha_p \vnu_p$

        \State $\ve_p \gets \ve_p + \alpha_p \ve_p^\nu$

            \State $\vgamma_p \gets \vchi \ve_p - \vj_p$

            \State $\vxi_p \gets \vf_p-\bm{\Phi}\vchi\ve_p$

            \State $\vg_{\vchi}^\old \gets \vg_{\vchi}$

            \State $\vg_{\vchi}$ is calculated by Eq.~\eqref{eq.gchi}

            \State $\vnu_{\vchi}\gets \vg_{\vchi}+\frac
                  {
                    \left\langle 
                      \vg_{\vchi},\vg_{\vchi}-\vg_{\vchi}^\old
                    \right\rangle
                    _{\mcD}
                  }
                  {
                      \left\|
                        \vg_{\vchi}^\old
                      \right\|
                      _{\mcD}^2
                  }
                  \vnu_{\vchi}$

            \State $\beta$ is determined by minimizing Eq.~\eqref{eq.chicost}

        \State $\vchi \gets \vchi+\beta \vnu_{\vchi}$

        \EndWhile\label{while}
        \State \textbf{return} $\vchi$
    \end{algorithmic}
  \end{algorithm}

  The CC-CSI method is given in Algorithm 1, where $\Re\{\cdot\}$ represents the real part operator and, correspondingly, the imaginary part operator is represented by $\Im\{\cdot\}$. Since $\mcM_{\mcD}$ always exists together with the stiffness matrix $\mA$, it is neglected for better readability in the remainder of this paper. 
  It is worth noting that $\vchi$ is assumed to be isotropic in this paper. Therefore, we average the two components of the contrast $\vchi$ for TE case, and the three components of the contrast $\vchi$ for 3-D case, after each update of $\vchi$.

\subsection{Initialization}\label{subsec.Initia}

  If no \textit{a priori} information about the objects is available, the contrast sources are initialized by (see \cite{habashy1994simultaneous,van1997contrast})
  \begin{equation}\label{eq.initvj}
    \vj_{p,0} = \frac{\|\bm{\Phi}^H\vf_p\|^2_{\mcD}}{\|\bm{\Phi}\bm{\Phi}^H\vf_p\|^2_{\mcS}}\bm{\Phi}^H\vf_p,
  \end{equation}
  which are obtained by back-propagation, multiplied by a weight to ensure that the data error is minimized. The contrast is initialized by (see \cite{van1999extended})
  \begin{equation}\label{eq.initchi}
    \vchi_0 = \left.{\displaystyle\sum_{p=1}^P\vj_{p,0}\overline{\ve_{p,0}}} \middle/ {\displaystyle\sum_{p=1}^P\ve_{p,0}\overline{\ve_{p,0}}}\right.,
  \end{equation}
  with $\ve_{p,0}=\ve_p^{\inc}+\mcM_{\mcD}\mA^{-1}\vj_{p,0}$.

\subsection{Computational complexity}\label{subsec.ComCom}
  
    In this subsection, we show that the CC-CSI method can be implemented without significantly increasing the computational complexity compared to the classical CSI method. Note that since the selecting matrix $\mcM_{\mcS}\in\mbC^{M\times N}$ has only $M<<N$ rows, the matrix $\bm{\Phi}$ can be calculated iteratively by solving $M$ linear systems of equations,
    \begin{equation}\label{eq.linsys}
        \mA^T\vvarphi_m={\mcM^m_{\mcS}}^T,\quad m=1,2,\dots,M,
    \end{equation}
    where, $\mcM^m_{\mcS}$ is the $m^{th}$ row of the selecting matrix $\mcM_{\mcS}$, and $(\cdot)^T$ represents the transpose operator. The matrix $\bm{\Phi}$ is assembled by $\bm\Phi = [\vvarphi_1,\vvarphi_2,\dots,\vvarphi_M]^T$. Since $\bm{\Phi}$ has only $M<<N$ rows, it is computationally much more efficient than the LU decomposition of the stiffness matrix $\mA$ (if we use LU decomposition). This feature makes it suitable to be computed and stored beforehand, which is of great importance for real applications, especially for 3-D inverse scattering problems. Although CC-CSI requires more matrix-vector multiplications, the extra computational cost is not significant by noting that the matrix $\bm{\Phi}$ has only $M<<N$ rows. This is further demonstrated in Subsection~\ref{subsec.ComTime}.

\section{Simulation results}\label{sec.SimRes}

\subsection{Configuration}\label{subsec.Configuration}

    In this section, the proposed algorithm is tested with a 2-D benchmark problem -- the ``Austria'' profile, which was also used in \cite{belkebir1996using,litman1998reconstruction,van2001contrast,van2003multiplicative}. Based on the benchmark problem, the performance of CC-CSI is analyzed in comparison to classical CSI and MR-CSI. 

\begin{figure}[!t]
        \centering
        \includegraphics[width=0.45\linewidth] {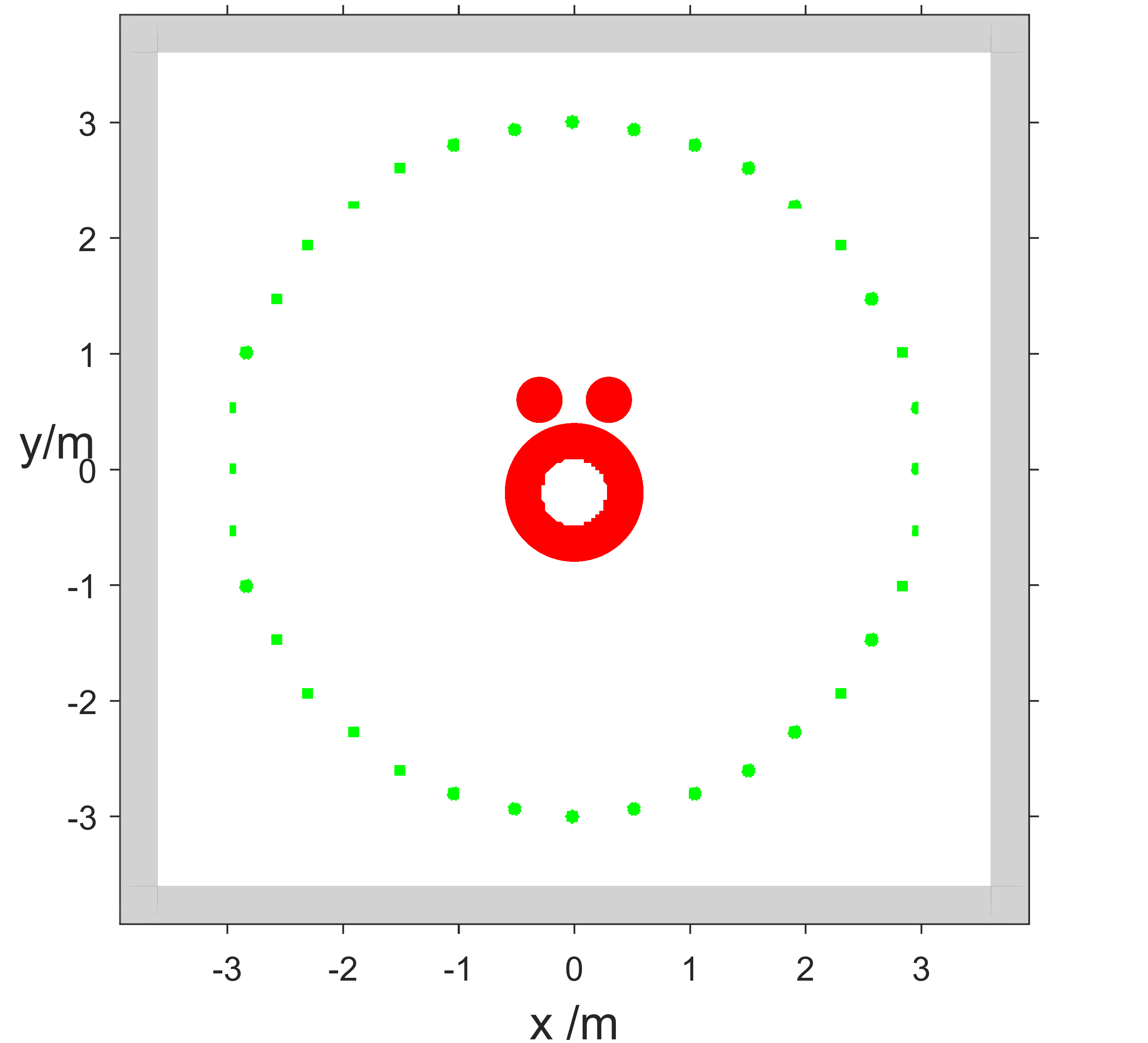}
        \caption{The original ``Austria'' profile contained in a region of $[-4,4]\times[-4,4]$ m$^2$. The green dots represent the 36 source/receiver stations. The boundaries of the four sides are terminated with PMLs. The two $z$-normal boundaries are subject to PBC.}
        \label{fig:configuration}
    \end{figure}

    Specifically, the objects to be inverted consist of two disks and one ring. Let us first establish our coordinate system such that the $z$-axis is parallel to the axis of the objects. The disks of radius 0.2 m are centred at ($-0.3$, 0.6) m and ($0.3$, 0.6) m. The ring is centred at (0,$-0.2$) m, and it has an exterior radius of 0.6 m and an inner radius of 0.3 m. Belkebir and Tijhuis \cite{belkebir1996using} and Litman et al. \cite{litman1998reconstruction} have used 64 sources and 65 receivers on a circle of radius 3 m centred at (0, 0), while the inverting domain was discretized into 30 $\times$ 30 cells. Van den Berg et al. \cite{van2001contrast,van2003multiplicative} have taken 48 source/receiver stations, while the inverting domain was discretized into 64 $\times$ 64 cells. In our simulation, 36 source/receiver stations are used and uniformly distributed on the same circle, which means we have under-sampled this problem further. In our simulation, $P=Q=36$, viz., we have $36\times36$ measurement data for TM case and $36\times72$ measurement data for TE case. Same objects but of different relative permittivity have been considered. The conductivity is fixed at 10 mS/m, so we have the same attenuation, while the relative permittivity attains values -- 2.0, 2.5, 3.0, and 3.5, respectively. The operating frequency is 300 MHz, therefore, the corresponding values of the contrast are $\chi=1.0-0.6\rmi$, $\chi=1.5-0.6\rmi$, $\chi=2.0-0.6\rmi$ and $\chi=2.5-0.6\rmi$, respectively. The original ``Austria'' profile is given in Fig.~\ref{fig:configuration}, in which the green dots represent the 36 source/receiver stations.

    The forward EM scattering problem is solved by a MATLAB-based 3-D FDFD package ``MaxwellFDFD'' \cite{maxwellfdfd-webpage}. The $x$- and $y$-normal boundaries are covered by perfect matching layers (PML) to simulate the anechoic chamber environment (see the gray layers of Fig.~\ref{fig:configuration} at the boundaries of the test domain), while the two $z$-normal boundaries are subject to periodic boundary conditions (PBC) to simulate the 2-D configuration. Line sources parallel to the $z$-axis are used to generate TM-polarized and TE-polarized incident wave. Non-uniform meshes are used to generate the scattered data, which means the testing domain is discretized with different mesh sizes determined by the distribution of the permittivity, viz., coarse meshes for low permittivity and fine meshes for high permittivity. The accuracy of the FDFD scheme is ensured by the following criterion \cite{W.Shin2013}
    \begin{equation}
        \Delta\leq\frac{\lambda_0}{15\sqrt{\varepsilon_r}},
    \end{equation} 
    where, $\lambda_0$ is the wavelength in free space. Non-uniform meshes greatly reduce the computational burden for solving the forward scattering problem. In contrast, uniform meshes are used to invert the scattered data, since we do not know the distribution of the permittivity beforehand. To guarantee the inverting accuracy, the following condition is satisfied
    \begin{equation}
        \Delta\leq\frac{\lambda_0}{15\sqrt{\max\{\varepsilon_r\}}}.  
    \end{equation} 
    The scattered field is obtained by subtracting the incident field from the total field. 

    \begin{figure}[!t]
        \centering
        \includegraphics[width=1.0\linewidth] {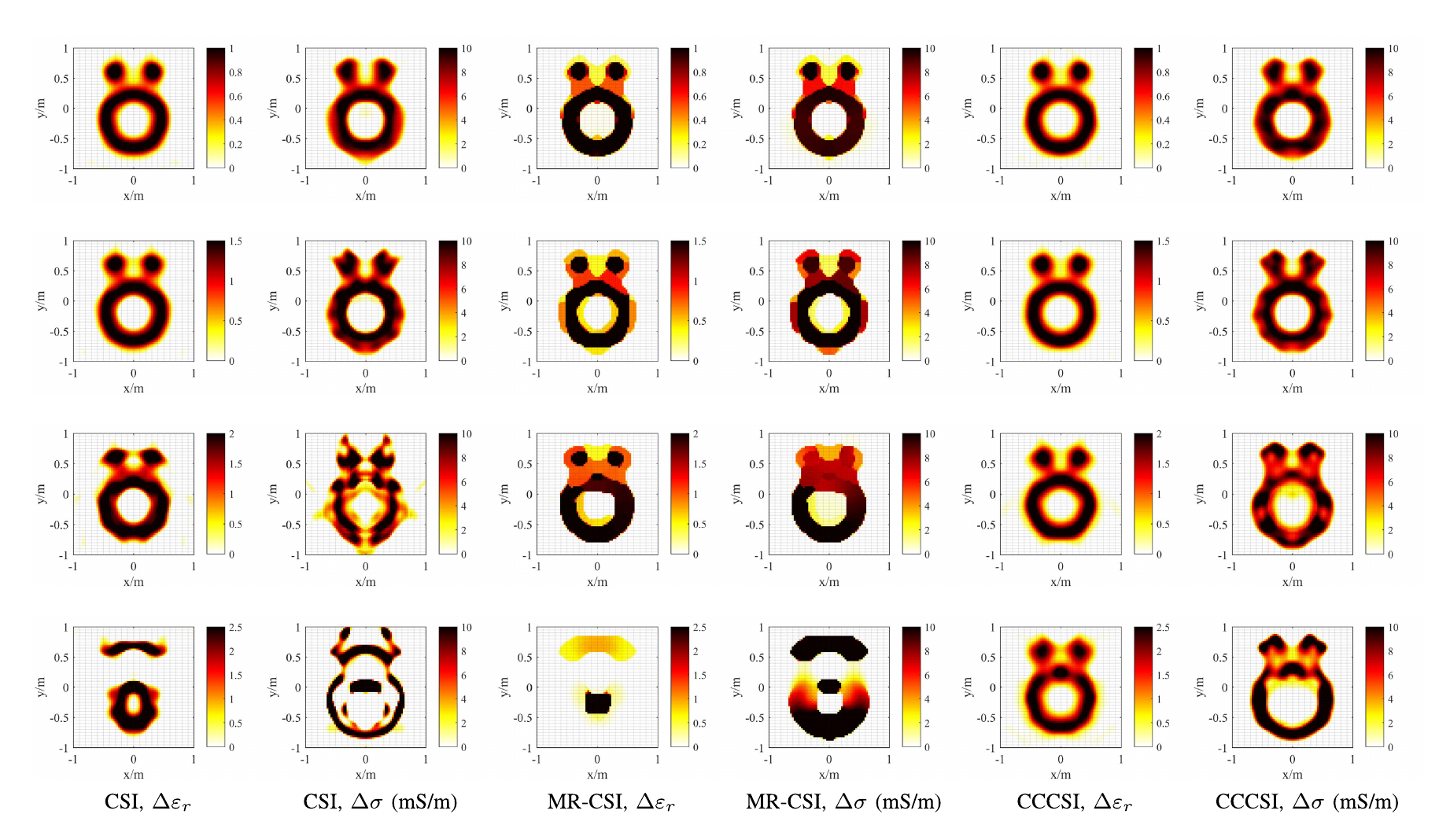}
        \caption{The relative permittivity and conductivity of the contrast obtained by classical CSI, MR-CSI, and CC-CSI, respectively, after 2048 iterations. The noise-free TM-polarized data at 300 MHz is processed. From top to bottom: $\chi=1.0-0.6\rmi$, $1.5-0.6\rmi$, $2.0-0.6\rmi$, and $2.5-0.6\rmi$.}
        \label{fig:TMInvNoiseFree}
    \end{figure}
  
\subsection{Condition number of the measurement matrix}\label{subsec.CondPhi}

    In the following simulations, the inversion domain is restricted to the region [-1.5,1.5] $\times$ [-1.5,1.5] m$^2$. The dimension of the mesh grid is 30 $\times$ 30 mm$^2$. Thus we have specifically in this simulation, $\bm{\Phi}\in\mbC^{36\times 10000}$ for TM polarization, and $\bm{\Phi}\in\mbC^{72\times 20000}$ for TE polarization. The condition numbers of $\bm{\Phi}$ are $\kappa(\bm{\Phi})_{TM}=5.19\times 10^2$ for TM polarization and $\kappa(\bm{\Phi})_{TE}=8.25\times 10^7$ for TE polarization. As one can see the condition number of the matrix $\bm{\Phi}$ is large for both TM and TE polarization, indicating that an error in the contrast sources $\vj_p$ may cause an increased error in the measurement data $\vf_p$. In addition, compared to TM polarization, TE polarization is more ill-conditioned because $\kappa(\bm{\Phi})_{TE}$ is much larger than $\kappa(\bm{\Phi})_{TM}$, due to the different formulation of the scattering equations. This implies that the introduction of the cross-correlated cost functional has higher influence on a TE case than a TM case, which is demonstrated by the following simulation results. It is worth noting that in the formulation of TE scattering problems, the operators involved have the same form, but one spatial dimension lower, compared to full 3-D scattering problems. Hence, the performance gain with CC-CSI in future 3-D inversion problems can be compared to the performance gain in the TE case.
\subsection{Noise-free data}\label{subsec.NoiseFree}

    \begin{figure}[!t]
        \centering
        \includegraphics[width=0.65\linewidth] {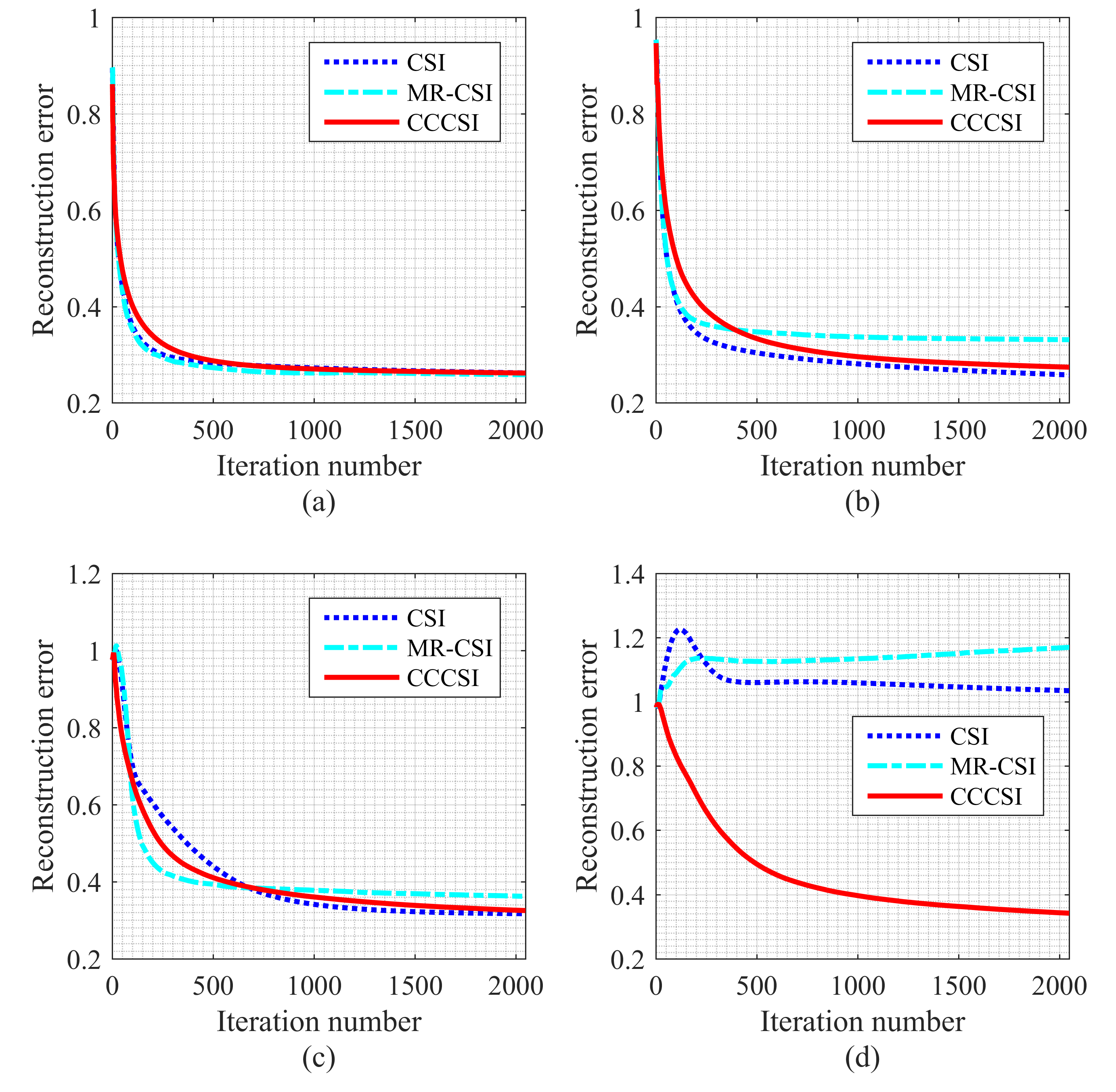}
        \caption{The reconstruction error curves of classical CSI, MR-CSI, and CC-CSI, in terms of the iteration number in the noise-free TM case. The operating frequency is 300 MHz. (a) $\chi=1.0-0.6\rmi$; (b) $\chi=1.5-0.6\rmi$; (c) $\chi=2.0-0.6\rmi$; (d) $\chi=2.5-0.6\rmi$.}
        \label{fig:TMErrNoiseFree}
    \end{figure}

\begin{figure}[!t]
        \centering
        \includegraphics[width=1.0\linewidth] {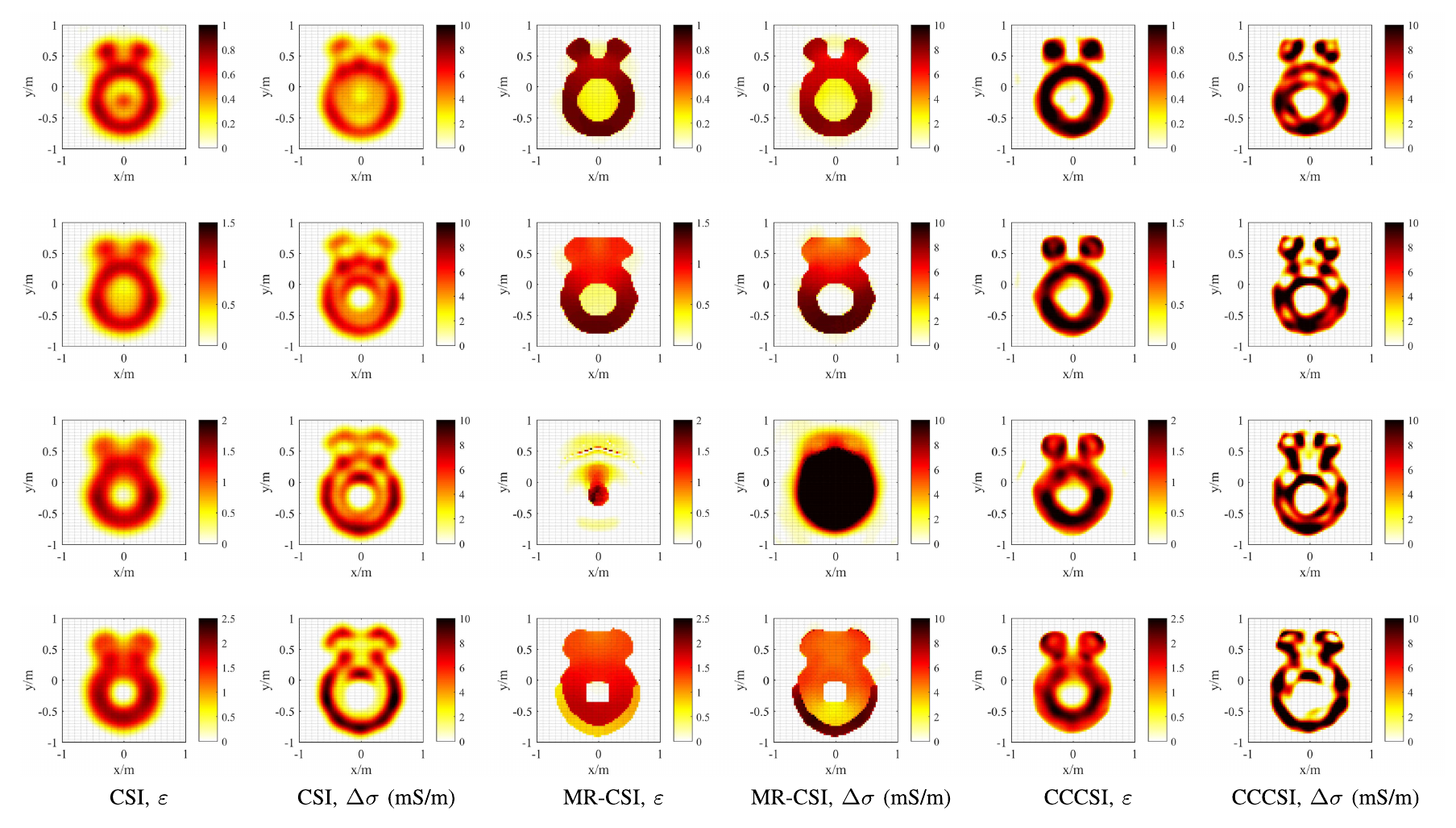}
        \caption{The relative permittivity and conductivity of the contrast obtained by classical CSI, MR-CSI, and CC-CSI, respectively, after 2048 iterations. The noise-free TE-polarized data at 300 MHz is processed. From top to bottom: $\chi=1.0-0.6\rmi$, $1.5-0.6\rmi$, $2.0-0.6\rmi$, and $2.5-0.6\rmi$.}
        \label{fig:TEInvNoiseFree}
    \end{figure}

    For fair comparison, in the following simulations, the contrast sources are initialized by Eq.~\eqref{eq.initvj} and the contrast is initialized by Eq.~\eqref{eq.initchi} for all the three algorithms. Since the background of this benchmark problem is free space, viz., $\Re\{\vchi\}\geq 0$ and $\Im\{\vchi\}\leq 0$, we exploit this \textit{a priori} information by simply enforcing the negative real part and the positive imaginary part of the contrast to zero after each update of the contrast \cite{van2001contrast}. Let us first investigate the inversion performance to the noise-free data. Both the TM-polarized data and the TE-polarized data are processed by classical CSI, MR-CSI, and CC-CSI, respectively. The relative permittivity and conductivity of the reconstructed contrast after 2048 iterations are shown in Fig.~\ref{fig:TMInvNoiseFree} for the TM case and Fig.~\ref{fig:TEInvNoiseFree} for the TE case. From Fig.~\ref{fig:TMInvNoiseFree} and Fig.~\ref{fig:TEInvNoiseFree} we see that MR-CSI generates blocky images because of the introduction of the total variation constraint, while classical CSI and CC-CSI have obvious variation in the reconstructed images. As we can see CC-CSI show better robustness by noting from Fig.~\ref{fig:TMInvNoiseFree} that the images of the contrast $\chi=2.0-0.6\rmi$ obtained by classical CSI and MR-CSI show more distortion than those of CC-CSI, and that classical CSI and MR-CSI fail to reconstruct the contrast of $\chi=2.5-0.6\rmi$ using the TM-polarized data. We can also see From Fig.~\ref{fig:TEInvNoiseFree} that MR-CSI fail to reconstruct the contrast $\chi=2.0-0.6\rmi$ using the noise-free TE-polarized data. In addition, we see that the interior hollow tube is better reconstructed by CC-CSI in the TE case, and the two smaller tubes are better distinguished by CC-CSI compared to classical CSI and MR-CSI.

\begin{figure}[!t]
        \centering
        \includegraphics[width=0.65\linewidth] {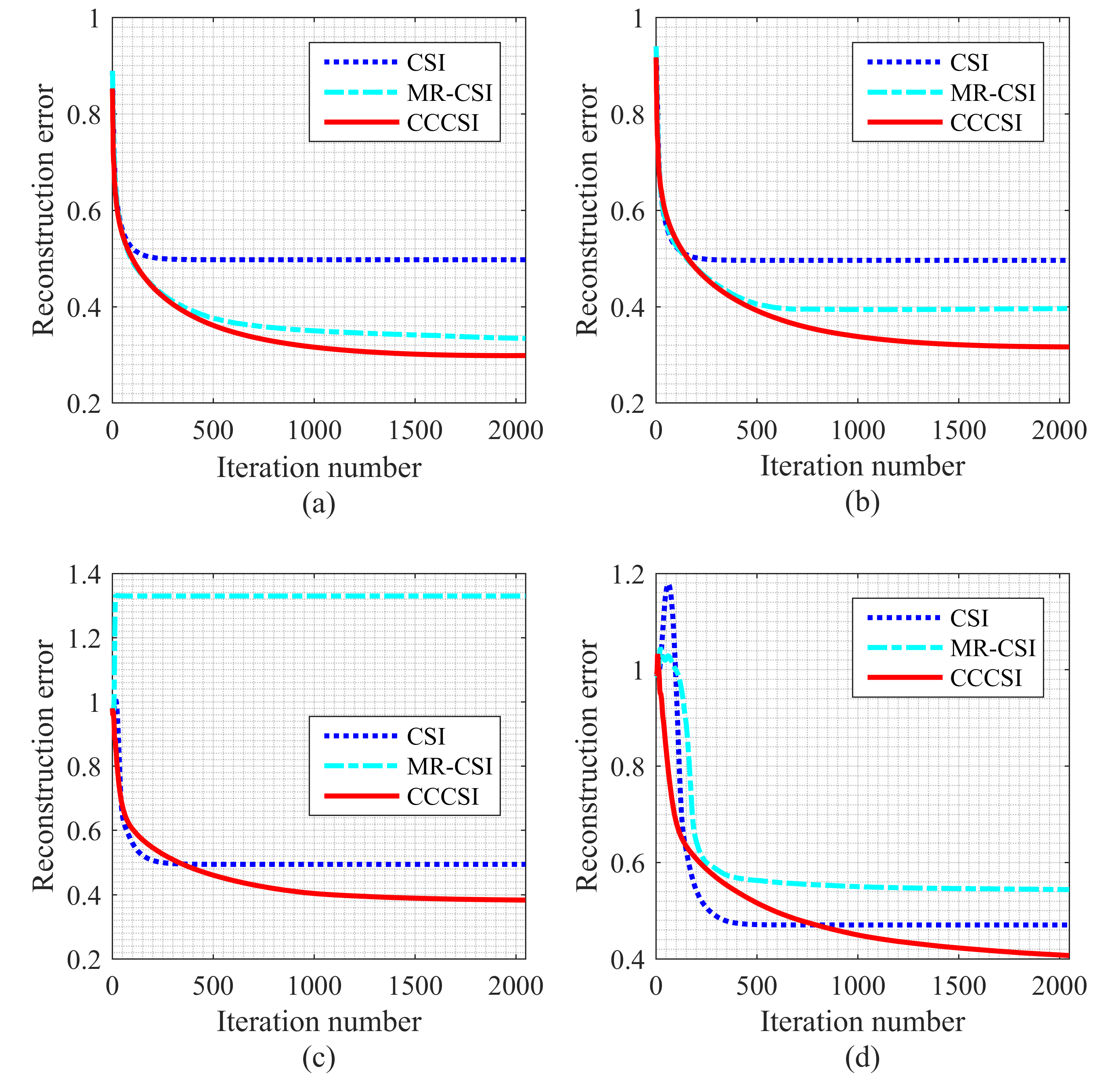}
        \caption{The reconstruction error curves of classical CSI, MR-CSI, and CC-CSI, in terms of the iteration number in the noise-free TE case. The operating frequency is 300 MHz. (a) $\chi=1.0-0.6\rmi$; (b) $\chi=1.5-0.6\rmi$; (c) $\chi=2.0-0.6\rmi$; (d) $\chi=2.5-0.6\rmi$.}
        \label{fig:TEErrNoiseFree}
    \end{figure}

    To quantitatively investigate the reconstruction accuracy, the reconstruction error of the three inversion methods is defined in the following as
    \begin{equation}
        \textit{err} = \frac{\|\vchi-\hat\vchi\|^2}{\|\vchi\|^2}.
    \end{equation}
    Fig.~\ref{fig:TMErrNoiseFree} and Fig.~\ref{fig:TEErrNoiseFree} give the comparison of the reconstruction error curves in terms of the iteration number of the three methods in the TM case and the TE case, respectively. As we can see quantitatively that, in the TM case, the three methods reach the same reconstruction errors in reconstructing the contrasts $\chi=1.0-0.6\rmi$, $1.5-0.6\rmi$, and $2.0-0.6\rmi$. However, the reconstruction errors of classical CSI and MR-CSI do not decrease in reconstructing the contrast $\chi=2.5-0.6\rmi$. In the contrast, the decreasing tendency of the reconstruction error curve of CC-CSI is not obviously affected by increasing the value of the contrast. In the TE case, we can see from Fig.~\ref{fig:TEErrNoiseFree} that CC-CSI can achieve lower reconstruction errors than classical CSI and MR-CSI, indicating the higher inversion accuracy of CC-CSI compared to classical CSI and MR-CSI. The reconstruction error of MR-CSI in reconstructing the contrast $\chi=2.0-0.6\rmi$ does not decrease, and the reconstruction error curve of classical CSI in reconstructing the contrast $\chi=2.5-0.6\rmi$ shows non-monotonicity. This demonstrates the poor robustness of both classical CSI and MR-CSI, and the better robustness of CC-CSI. 

    It is worth noting that CC-CSI shows higher inversion accuracy than classical CSI and MR-CSI in the TE case, but similar inversion accuracy with classical CSI and MR-CSI in TM case. Recalling the previous subsection, we know that the matrix $\bm{\Phi}$ of TE polarization has a much larger condition number than that of TM polarization. Therefore, compared to the TM case, same level of cross-correlated error presented in the measurement domain in TE case corresponds to a smaller reconstruction error in the field domain.   

    In addition, as we can see from the simulation results, MR-CSI shows not only poor robustness, but also unstable performance with respect to inversion accuracy compared to classical CSI and CC-CSI. As is well known, total variation regularization was originally proposed for noise removal in the digital image processing \cite{rudin1992nonlinear}. Obviously, the feasibility condition for applying the total variation regularization is that this noisy image is suitable for processing. However, this is apparently not the case in CSI, because the image of the contrast in CSI is optimized iteratively. In the design of MR-CSI, total variation constraint is very likely to be applied to a seriously distorted image of the contrast in the beginning, and thus may mislead and degrade the optimization process. Therefore, benefits can be possibly obtained from MR-CSI only if the contrast can be reliably reconstructed with CSI. Namely, the benefits from MR-CSI are not guaranteed. This perfectly explains the instable performance of MR-CSI shown in Fig.~\ref{fig:TMErrNoiseFree} and Fig.~\ref{fig:TEErrNoiseFree}. 

\subsection{Noise-disturbed data}\label{subsec.Noise}

    \begin{figure}[!t]
        \centering
        \includegraphics[width=1.0\linewidth] {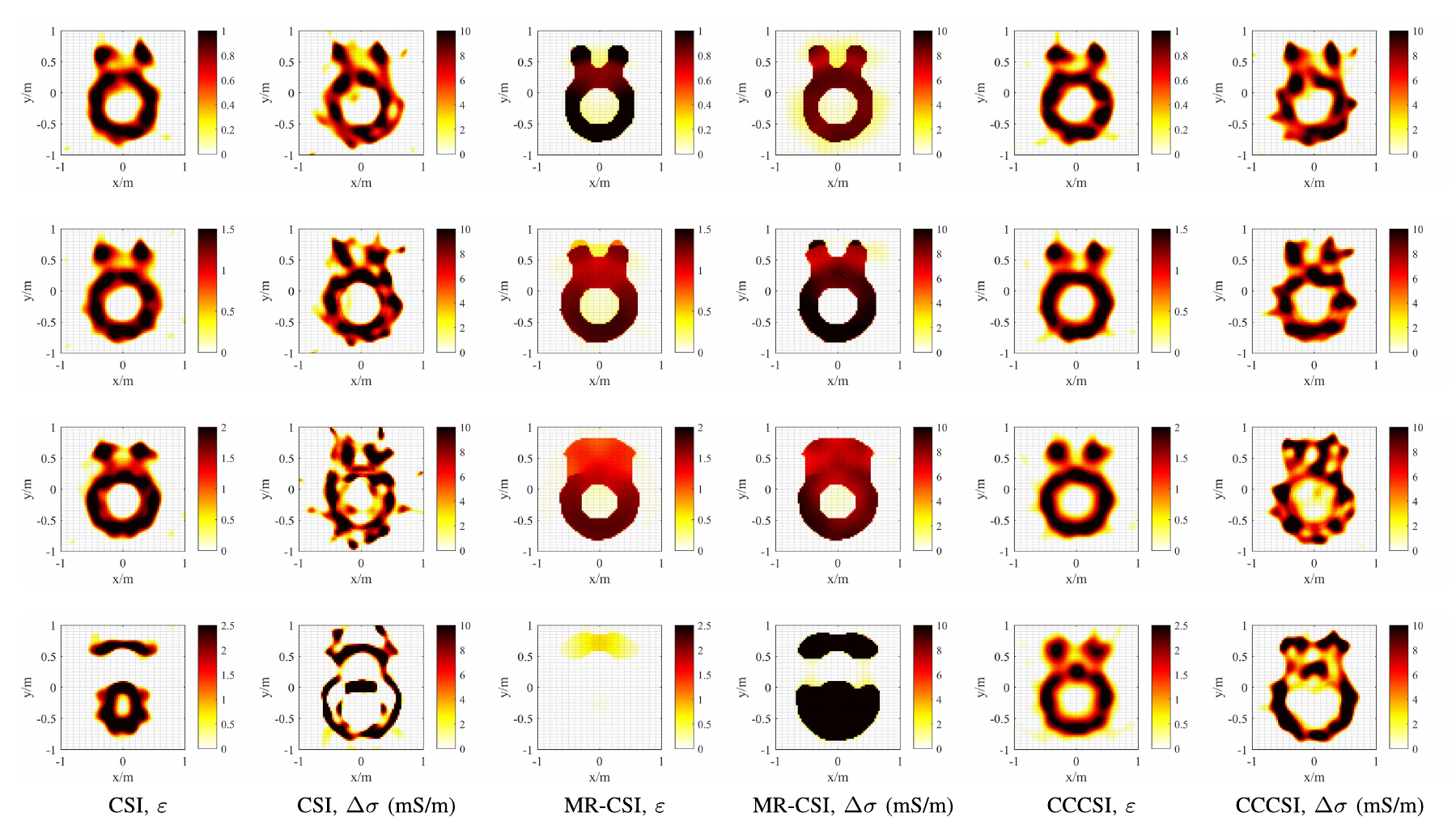}
        \caption{The relative permittivity and conductivity of the contrast obtained by classical CSI, MR-CSI, and CC-CSI, respectively, after 2048 iterations. The noise-disturbed TM-polarized data at 300 MHz is processed. From top to bottom: $\chi=1.0-0.6\rmi$, $1.5-0.6\rmi$, $2.0-0.6\rmi$, and $2.5-0.6\rmi$. 10\% additive random white noise is considered.}
        \label{fig:TMInv}
    \end{figure}

    \begin{figure}[!t]
        \centering
        \includegraphics[width=0.65\linewidth] {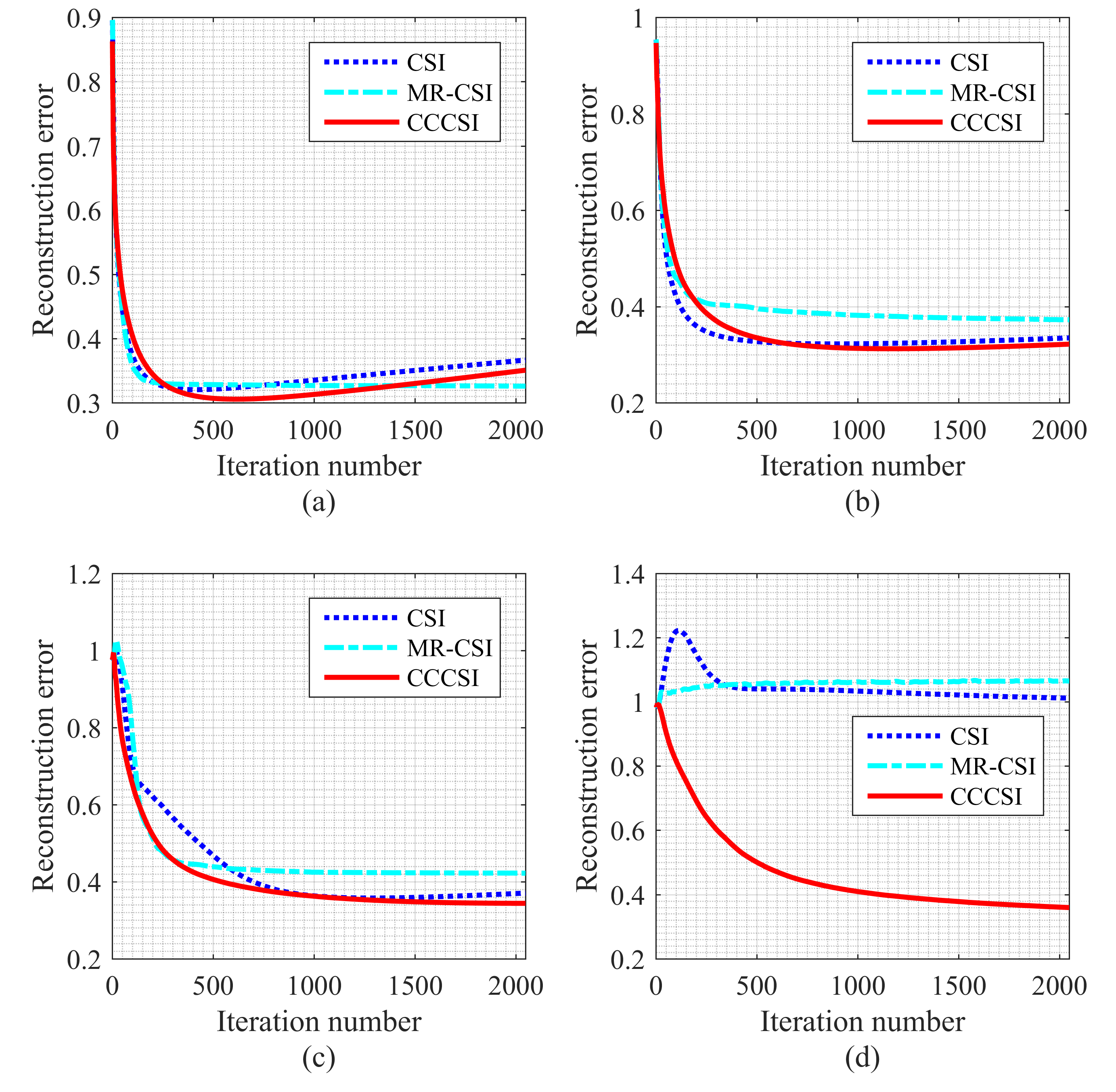}
        \caption{The reconstruction error curves of classical CSI, MR-CSI, and CC-CSI, in terms of the iteration number in the noise-disturbed TM case. The operating frequency is 300 MHz. 10\% additive random white noise is considered. (a) $\chi=1.0-0.6\rmi$; (b) $\chi=1.5-0.6\rmi$; (c) $\chi=2.0-0.6\rmi$; (d) $\chi=2.5-0.6\rmi$.}
        \label{fig:TMErr}
    \end{figure}

    \begin{figure}[!t]
        \centering
        \includegraphics[width=1.0\linewidth] {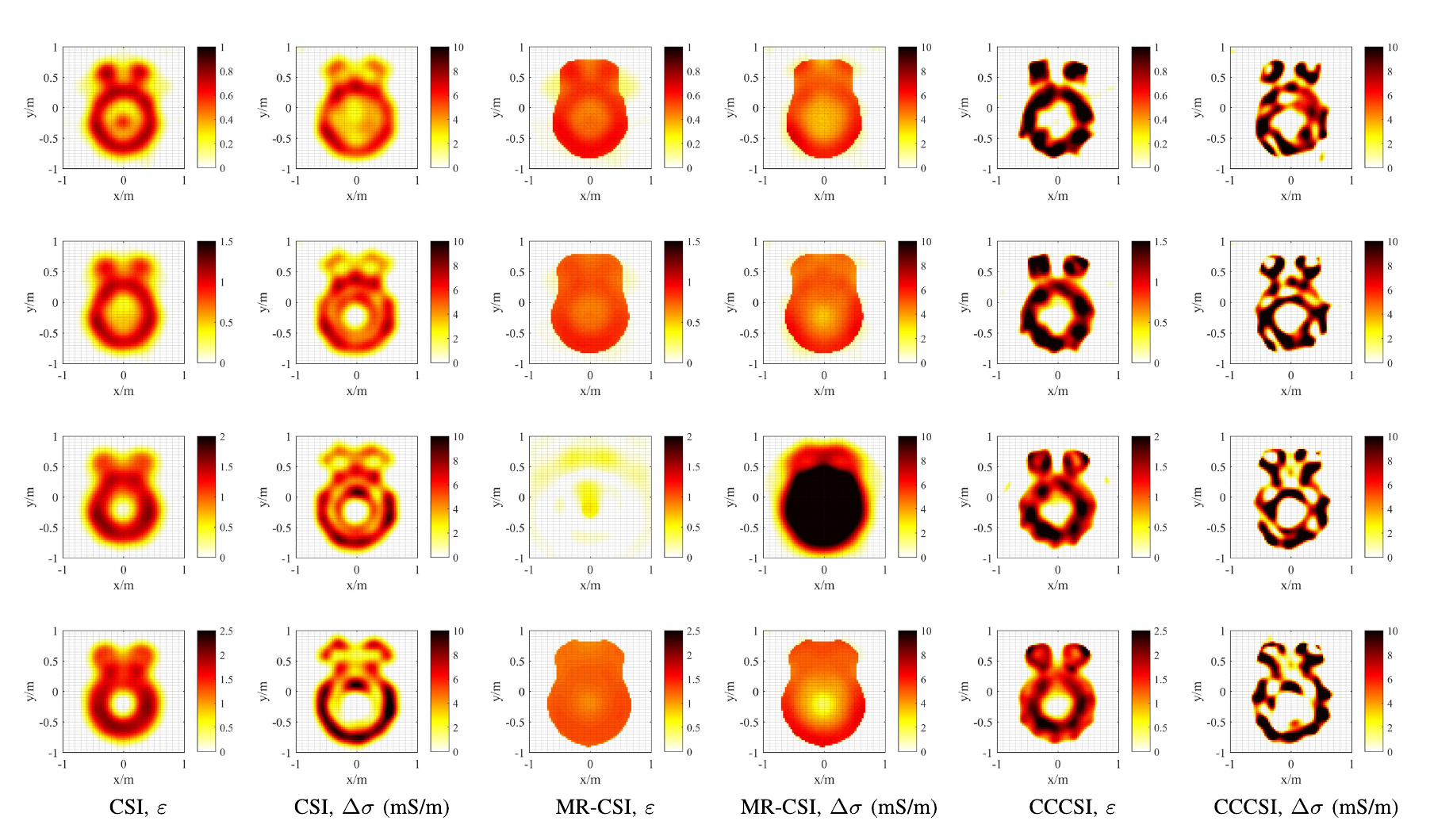}
        \caption{The relative permittivity and conductivity of the contrast obtained by classical CSI, MR-CSI, and CC-CSI, respectively, after 2048 iterations. The noise-disturbed TE-polarized data at 300 MHz is processed. From top to bottom: $\chi=1.0-0.6\rmi$, $1.5-0.6\rmi$, $2.0-0.6\rmi$, and $2.5-0.6\rmi$. 10\% additive random white noise is considered.}
        \label{fig:TEInv}
    \end{figure}

    \begin{figure}[!t]
        \centering
        \includegraphics[width=0.65\linewidth] {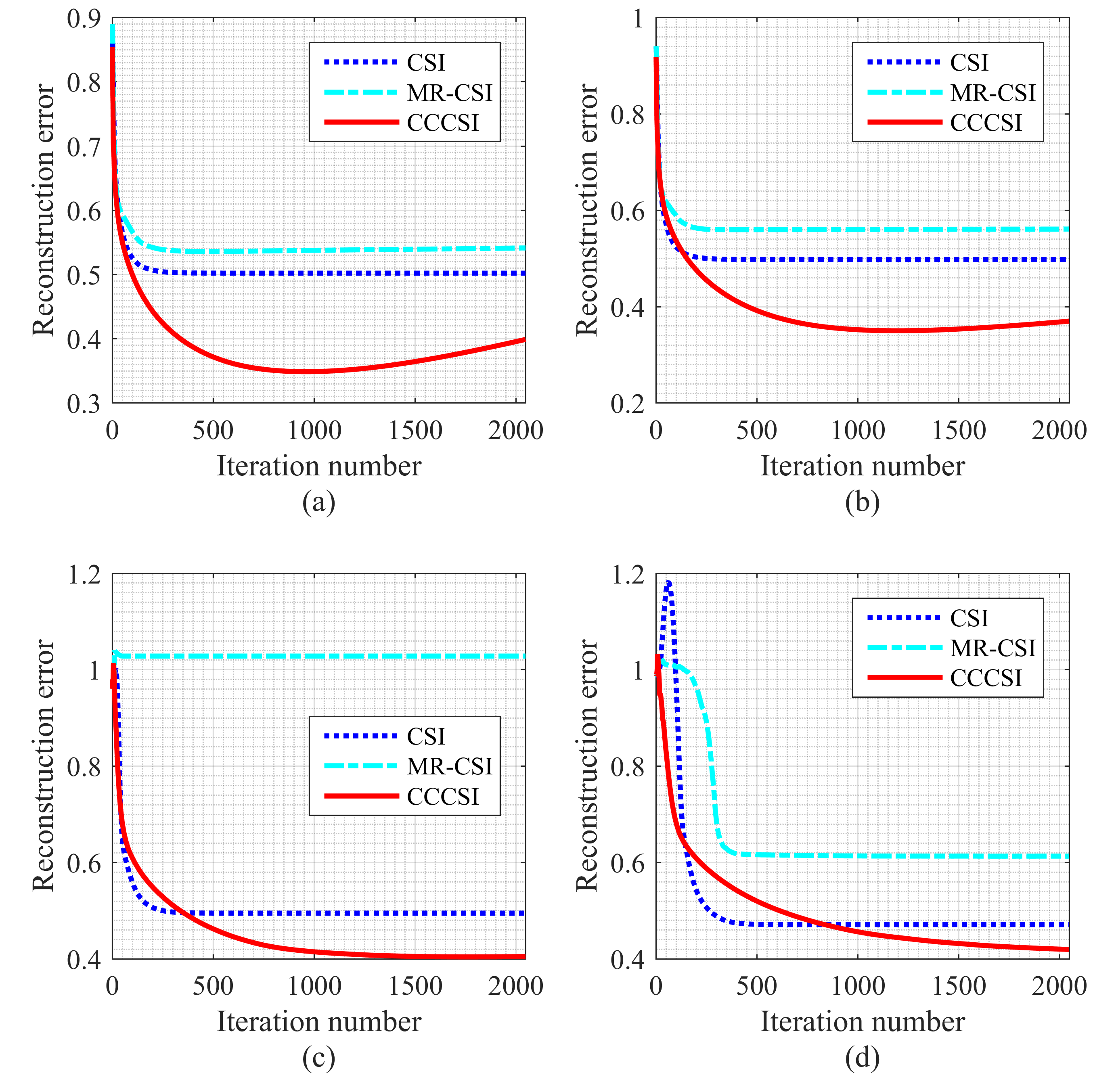}
        \caption{The reconstruction error curves of classical CSI, MR-CSI, and CC-CSI, in terms of the iteration number in the noise-disturbed TE case. The operating frequency is 300 MHz. 10\% additive random white noise is considered. (a) $\chi=1.0-0.6\rmi$; (b) $\chi=1.5-0.6\rmi$; (c) $\chi=2.0-0.6\rmi$; (d) $\chi=2.5-0.6\rmi$.}
        \label{fig:TEErr}
    \end{figure}

    In real applications, the measurement data are very likely to be disturbed by noises. Apart from that, there is always error in modeling the incident fields. In this subsection, we investigate the inversion performance of the three methods to noise-disturbed data while the incident fields are assumed to be exactly known. Random white noise is added to the measurement data following the same procedure used in \cite{van2003multiplicative}, 
    \begin{equation}
      \vf_{p,noise} = \vf_p+\zeta\times\underset{m}{\max}\{|f_{p,m}|\}(\vn_1+\rmi\vn_2),
    \end{equation}
    with $p = 1,2,\cdots,P$, $m = 1,2,\cdots,M$. Here, $\vn_1$ and $\vn_2$ are two random numbers varying from $-1$ up to 1, $\zeta$ = 10\% is the amount of noise, and $\underset{m}{\max}\{|f_{p,m}|\}$ gives the largest value among the amplitudes of the $M$ measurement data, which means the noise is scaled by the largest amplitude of the measurement data. $M=36$ in the TM case and $M=72$ in the TE case.

    Fig.~\ref{fig:TMInv} and Fig.~\ref{fig:TEInv} show the inverted results after 2048 iterations by the three methods using the TM-polarized data and the TE-polarized data, respectively. In comparison to Fig.~\ref{fig:TMInvNoiseFree} and Fig.~\ref{fig:TEInvNoiseFree}, we can see obvious distortion in the reconstructed images because of the disturbance by the additive random noise. What's worse, we see from Fig.~\ref{fig:TEInv} that the hollow tube is not distinguishable any more in the images obtained by MR-CSI using the noise-disturbed TE-polarized data. As mentioned previously in Subsection~\ref{subsec.NoiseFree}, since the hollow tube cannot be well recognized by CSI (see the images obtained by CSI in Fig.~\ref{fig:TEInv}), we lose the basic feasibility condition for applying the total variation constraint and therefore the hollow tube is reconstructed by MR-CSI to a solid one. In the contrast, CC-CSI is still capable to distinguish the hollow tube even though the measurement data has been disturbed by 10\% additive random noise.

    The reconstruction error curves of classical CSI, MR-CSI and CC-CSI in the noise-disturbed cases are given in Fig.~\ref{fig:TMErr} for the TM case and Fig.~\ref{fig:TEErr} for the TE case. As we can see, the error curves of classical CSI and MR-CSI show instability as the contrast goes higher, while the error curves of CC-CSI are always monotonously decreasing, indicating again the better robustness of CC-CSI. We can also see from Fig.~\ref{fig:TEErr} that CC-CSI has lower reconstruction error than classical CSI and MR-CSI, which is consistent with the simulation results in the noise-free cases. 

    What's more, from Fig.~\ref{fig:TMErr}(a) we see that the reconstruction errors of classical CSI and CC-CSI turn out to increase as the iteration goes on. Same phenomenon also occurs in Fig.~\ref{fig:TEErr}(a,b) for CC-CSI. By comparison to Fig.~\ref{fig:TMErrNoiseFree}(a) and Fig.~\ref{fig:TEErrNoiseFree}(a,b) in the noise-free case, this phenomenon can been easily explained by the introduction of the additive random noise. Therefore, in real applications, a good termination condition is critical for preventing the methods from over-fitting the noise and for saving computation time.

\subsection{Computational performance}\label{subsec.ComTime}

  As mentioned previously, CC-CSI can be implemented without significantly increasing the computational complexity. To demonstrate this point, we ran the MATLAB codes on a desktop with one Intel(R) Core(TM) i5-3470 CPU @ 3.20 GHz, and we did not use parallel computing. The computation times of classical CSI, MR-CSI, and CC-CSI, running for 2048 iterations are given in Table~\ref{tab.TM}. 
  
  \begin{table}[!ht]
  \renewcommand{\arraystretch}{1.3}
  \caption{The computational times}
  \label{tab.TM}
  \centering 
  \begin{tabular}{|c|c|c|c|}
  \hline
  \diagbox{Polarization}{Time /s}{Methods} & CSI & MR-CSI & CC-CSI\\
  \hline
  TM & 1361.2 & 1419.7 & 1429.6 \\ \hline
  TE & 2931.9 & 3040.9 & 3146.2 \\ \hline
  \end{tabular}
  \end{table}
  
  As we can see CSI is the most efficient, MR-CSI is in the middle, and CC-CSI runs slightly longer. If we define the increment percentage of the running times of CC-CSI as
  \begin{equation}
    \frac{T_\CCCSI-T_\CSI}{T_\CSI}\times 100\%,
  \end{equation}
  we have that the running time of CC-CSI is slightly longer than that of classical CSI by around 5.0\% in the TM case and 7.3\% in the TE case.

\section{Conclusion}\label{sec.conclusion}

  In this paper, a cross-correlated contrast source inversion (CC-CSI) method is proposed by modifying the cost functional of the CSI method to interrelate the state error and the data error. The proposed algorithm is tested with a 2-D benchmark problem which has also been tested by Belkebir and Tijhuis \cite{belkebir1996using}, Litman et al. \cite{litman1998reconstruction}, and Berg et al. \cite{van2001contrast,van2003multiplicative}. The simulation results with both TM-polarized wave and TE-polarized wave show that CC-CSI outperforms classical CSI and MR-CSI with respect to robustness and inversion accuracy, especially in the TE case. Which shows to be promising for the robustness and inversion accuracy in full 3-D inversion problems. We have also shown that CC-CSI can be implemented without significantly increasing the computational burden. As the Maxwell equations are formulated within a 3-D finite difference frequency domain (FDFD) scheme, it is straightforward to extend the proposed inversion scheme to future 3-D inverse scattering problems. Numerical results of 3-D scattering objects, including the application of the proposed method to experimental data will be published in future work.

\appendix

\section{Derivation of the step size \texorpdfstring{$\alpha_{p,n}$}{LG}}\label{sec.apdixA}
  First, let us rewrite the cost function $C_\CCCSI^{\vj}(\vchi_{n-1}, \vj_{p,n-1} + \alpha_{p}\vnu_{p,n})$ as follows
  \begin{equation}
  \begin{split}
    &C_\CCCSI^{\vj}(\vchi_{n-1}, \vj_{p,n-1} + \alpha_{p}\vnu_{p,n}) = \eta^{\mcD}_{n-1}\sum_{p=1}^P\left\|\vgamma_{p,n-1}+\alpha_{p}(\vchi_{n-1}\mcM_{\mcD}\mA^{-1}-\bm{\mI})\vnu_{p,n}\right\|^2 + \\
    &\eta^{\mcS}\sum_{p=1}^P\left\|\vrho_{p,n-1}-\alpha_{p}\bm{\Phi}\vnu_{p,n}\right\|^2 + \eta^{\mcS}\sum_{p=1}^P\left\|\vxi_{p,n-1}-\alpha_{p}\bm{\Phi}\vchi_{n-1}\mcM_{\mcD}\mA^{-1}\vnu_{p,n}\right\|^2
  \end{split}
  \end{equation}
  Obviously, it can be further simplified in the form of
  \begin{equation}
  \begin{split}
    C_\CCCSI^{\vj}(\vchi_{n-1}, \vj_{p,n-1} + \alpha_{p}\vnu_{p,n}) = &a_{p,2}\alpha_{p,n}^2+a_{p,1}\alpha_{p,n}+a_{p,0}+b_{p,2}\alpha_{p}^2+b_{p,1}\alpha_{p}+b_{p,0}+\\
    &c_{p,2}\alpha_{p,n}^2+c_{p,1}\alpha_{p}+c_{p,0}.
  \end{split}
  \end{equation}
  Therefore, we have
  \begin{equation}
    \alpha_{p,n} = \underset{\alpha_p}{\max}\ C_\CCCSI^{\vj}(\vchi_{n-1}, \vj_{p,n-1} + \alpha_{p}\vnu_{p,n}) = -\frac12\frac{a_{p,1}+b_{p,1}+c_{p,1}}{a_{p,2}+b_{p,2}+c_{p,2}}.
  \end{equation}
  Note that 
  \begin{subequations}\label{eq.pars1}
  \begin{align}
    a_{p,2} &= \eta^{\mcS}\left\|\bm{\Phi}\vnu_{p,n}\right\|_{\mcS}^2,\label{eq.pars_a2}\\
    a_{p,1} &= -2\eta^\mcS \Re\left\{\vnu_{p,n}^H\bm{\Phi}^H\vrho_{p,n-1}\right\},\label{eq.pars_a1}
  \end{align}
  \end{subequations}
  \begin{subequations}\label{eq.pars2}
  \begin{align}
    b_{p,2} &= \eta_{n-1}^\mcD\|\vnu_{p,n}-\vchi \mcM_{\mcD}\mA^{-1}\vnu_{p,n}\|_\mcD^2,\label{eq.pars_b2}\\
    b_{p,1} &= 2\eta^\mcD_{n-1} \Re\left\{\vnu_{p,n}^H(\vchi\mcM_{\mcD}\mA^{-1}-\mI)^H\vgamma_{p,n-1}\right\},\label{eq.pars_b1}
  \end{align}
  \end{subequations}
  \begin{subequations}\label{eq.pars3}
  \begin{align}
    c_{p,2} &= \eta^{\mcS}\left\|\bm{\Phi}\vchi\mcM_{\mcD}\mA^{-1}\vnu_{p,n}\right\|_{\mcS}^2,\label{eq.pars_c2}\\
    c_{p,1} &= -2\eta^\mcS \Re\left\{\vnu_{p,n}^H(\bm{\Phi}\vchi\mcM_{\mcD}\mA^{-1})^H\vxi_{p,n-1}\right\},\label{eq.pars_c1}
  \end{align}
  \end{subequations}
  and 
  \begin{equation}
  \begin{split}
    \vg_{p,n} = &-2\eta^\mcS\bm{\Phi}^H\vrho_{p,n-1} + 2\eta^\mcD_{n-1}(\vchi\mcM_{\mcD}\mA^{-1}-\mI)^H\vgamma_{p,n-1} - 2\eta^\mcS(\bm{\Phi}\vchi\mcM_{\mcD}\mA^{-1})^H\vxi_{p,n-1},
  \end{split}
  \end{equation}
  it is easy to obtain that 
  \begin{equation}\label{eq.alpha}
    \alpha_{p,n} = -\frac{\Re\left\{\left\langle \vg_{p,n},\vnu_{p,n}\right\rangle_\mcD\right\}}
                  {2(a_{p,2}+b_{p,2}+c_{p,2})}.
  \end{equation} 
  where, $a_{p,2}$, $b_{p,2}$, and $c_{p,2}$ are given by Eq.~\eqref{eq.pars_a2}, Eq.~\eqref{eq.pars_b2}, and Eq.~\eqref{eq.pars_c2}, respectively.




%
\bibliographystyle{ieeetr}
\bibliography{mybib.bib}

\end{document}